\documentclass[5p, preprint, times]{elsarticle}
\makeatletter
\def\ps@pprintTitle{%
 \let\@oddhead\@empty
 \let\@evenhead\@empty
 \def\@oddfoot{}%
 \let\@evenfoot\@oddfoot}
\makeatother

\usepackage{units}
\usepackage[german, british]{babel}
\usepackage{setspace}
\usepackage[section]{placeins}
\usepackage[hang]{subfigure}
\usepackage{tabularx}
\usepackage{amsmath}
\usepackage{amssymb}
\usepackage[figuresright]{rotating}

\begin{document}

\begin{frontmatter}

\title{Low-Temperature Relative Reflectivity Measurements of Reflective and Scintillating Foils used in Rare Event Searches}

\cortext[cor1]{Corresponding author}

\author[TUM]{A. Langenk\"amper\corref{cor1}}
\ead{alexander.langenkaemper@tum.de}
\author[TUM]{A. Ulrich\corref{cor1}}
\ead{andreas.ulrich@ph.tum.de}
\author[Cluster]{X. Defay}
\author[TUM]{F. v. Feilitzsch}
\author[TUM]{J.-C. Lanfranchi}
\author[TUM]{E. Mondrag\'on}
\author[TUM]{A. M\"unster}
\author[TUM]{C. Oppenheimer}
\author[TUM]{W. Potzel}
\author[TUM]{S. Roth}
\author[TUM]{S. Sch\"onert}
\author[TUM]{H. Steiger}
\author[TUM]{H. H. Trinh Thi}
\author[TUM]{S. Wawoczny}
\author[TUM]{M. Willers}
\author[TUM]{A. Z\"oller}

\address[TUM]{Physik-Department, Technische Universtit\"at M\"unchen, D-85748 Garching, Germany}
\address[Cluster]{Excellence Cluster Universe, Technische Universit\"at M\"unchen, D-85748 Garching, Germany}

\begin{abstract}
In this work we investigate the reflectivity of highly reflective multilayer polymer foils used in the CRESST experiment. The CRESST experiment searches directly for dark matter via operating scintillating CaWO$_4$ crystals as targets for elastic dark matter-nucleon scattering. In order to suppress background events, the experiment employs the so-called phonon-light technique which is based on the simultaneous measurement of the heat signal in the main CaWO$_4$ target crystal and of the emitted scintillation light with a separate cryogenic light detector. Both detectors are surrounded by a highly reflective and scintillating multilayer polymer foil to increase the light collection efficiency and to veto surface backgrounds. While this study is motivated by the CRESST experiment, the results are also relevant for other rare event searches using scintillating cryogenic bolometers in the field of the search of dark matter and neutrinoless double beta decay ($0\nu\beta\beta$). In this work a dedicated experiment has been set up to determine the relative reflectivity at $\unit[300]{K}$ and $\unit[20]{K}$ of three multilayer foils (``VM2000'', ``VM2002'', ``Vikuiti'') produced by the company 3M. The intensity of a light beam reflected off the foil is measured with a CCD camera. The ratio of the intensities at $\unit[300]{K}$ and $\unit[20]{K}$ corresponds to the relative reflectivity change. The measurements performed in this work show no variation of the reflectivity with temperature at a level of $\sim\unit[1]{\%}$.
\end{abstract}


\end{frontmatter}

\section{Introduction}
\label{Intro}

The direct dark matter search experiment CRESST\footnote{Cryogenic Rare Event Search with Superconducting Thermometers} aims at the detection of dark matter particles elastically scattering off target nuclei in scintillating CaWO$_4$ single crystals which are operated as cryogenic detectors at mK temperatures \cite{Angloher2016, Angloher2014, Angloher2009, Angloher2004}.\\
For active background discrimination CRESST is using the so-called phonon-light technique based on the simultaneous measurement of the heat signal in a cryogenic phonon detector (CaWO$_4$ crystal \cite{Erb2013, Muenster2014} equipped with a thermometer) and a light signal with a separate cryogenic light detector (Silicon-on-Sapphire plate equipped with thermometer) \cite{Angloher2004}. A particle interaction in the crystal leads to an energy deposition. This energy is primarily transformed into phonons (heat) and a few percent into photons, depending on the type of particle interaction \cite{Angloher2016}. Both detectors are enclosed in a reflective and scintillating housing which consists of a highly reflecting multilayer polymer foil to increase the light collection and veto surface backgrounds.\\
In the CRESST experiment, only a fraction of the emitted scintillation light is detected due to a limited transparency of the target crystal as well as the size and absorptivity of the light detector. Furthermore, a decreased light collection could be explained due to the possible limitations of the reflecting and scintillating foil since the reflectivity is only measured at room temperature \cite{Kiefer2016, Janecek1, Janecek2, Motta}. The foils used in CRESST are multilayer polymer mirrors and have a high reflectivity of $>\unit[98]{\%}$ in the wavelength range from 400 to $\unit[800]{nm}$ at $\unit[300]{K}$ \cite{3M}. Due to the layered structure of the foils under investigation \cite{ScienceFoils}, a potential contraction of the layers could lead to a change in the interference between the different layers and therefore, a decreased reflectivity at low temperatures cannot be excluded. To investigate the temperature dependence, relative measurements of the reflectivity at temperatures of $T=\unit[300]{K}$ and $T=\unit[20]{K}$ have been performed. The temperature of $\unit[20]{K}$ represents the lowest temperature accessible with the cryocooler used. The $\unit[20]{K}$ missing to the mK base temperature of \cite{Angloher2016, Angloher2014, Angloher2009, Angloher2004} represent only $\sim\unit[8]{\%}$ of the range from room- to the base temperature and a potential change of the reflectivity should appear up to the temperature of $\unit[20]{K}$. For this purpose, a dedicated setup was built in the framework of this study to investigate three foils (``Vikuiti'', ``VM2000'' and ``VM2002'' (3M)). \\

\section{Experimental Setup}
\label{Setup}

\begin{figure} [htbp]
\centering
{\includegraphics[width=0.4\textwidth]{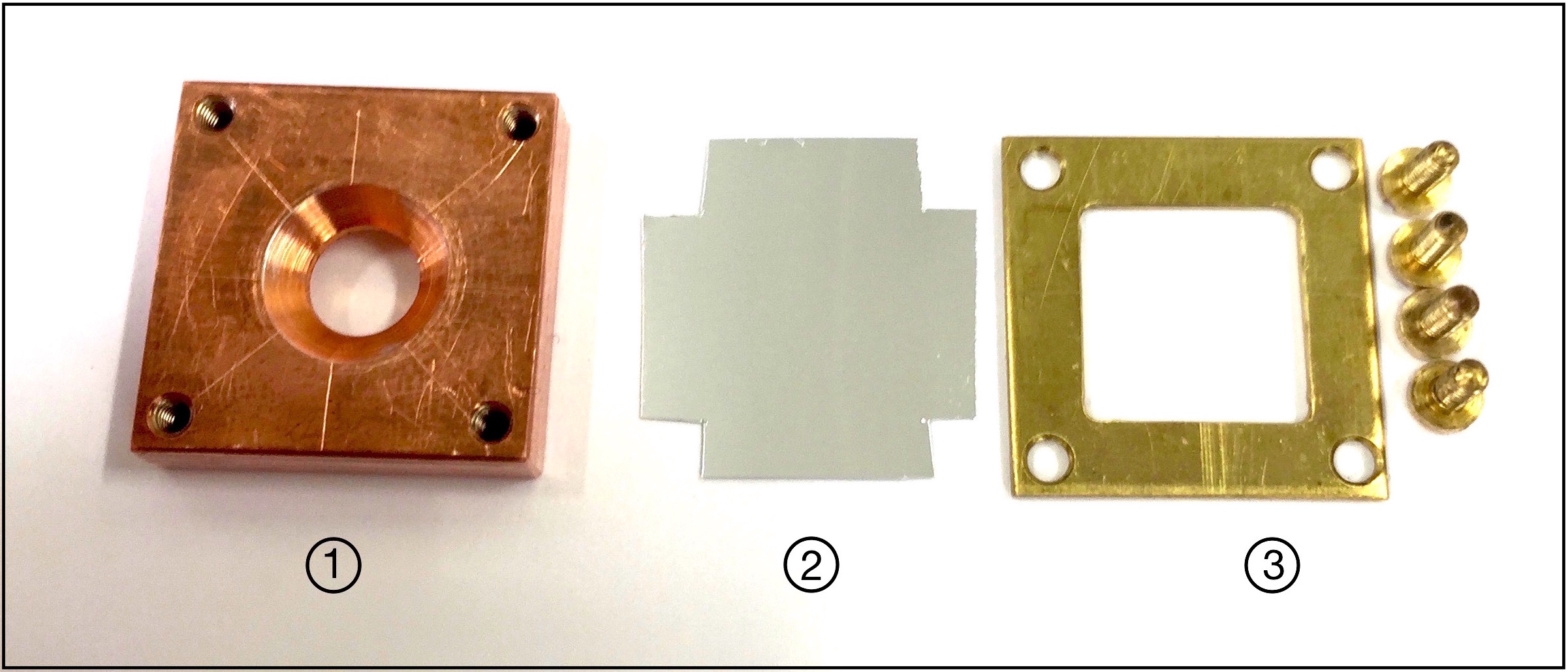}}
\caption{Holder for the foil samples. A small amount of vacuum grease is put onto the upper surface of a $\unit[2\text{\,x\,}2]{cm^2}$ copper plate (1) as well as onto the brass frame (3) to allow movements of the foil. The foil (2) is fixed in between these two parts which are screwed together.} 
\label{Holder}
\end{figure}

The concept of the experiments is based on a comparison of the measured intensities of a light beam which is reflected off a foil sample at $\unit[300]{K}$ and at $\unit[20]{K}$. From the ratio of these intensities the relative change in reflectivity can be derived. The advantage of this setup is that it is not sensitive to changes of the reflection due to thermal contraction of the holder and the foil and therefore, is not sensitive to any image distortion up to the size of the 2D sensor used. By integrating the signal over the whole area of the sensor it is possible to compare the results of the measurements at $\unit[300]{K}$ and $\unit[20]{K}$.\\
\begin{figure} [htbp]
\centering
\includegraphics[width=0.36\textwidth]{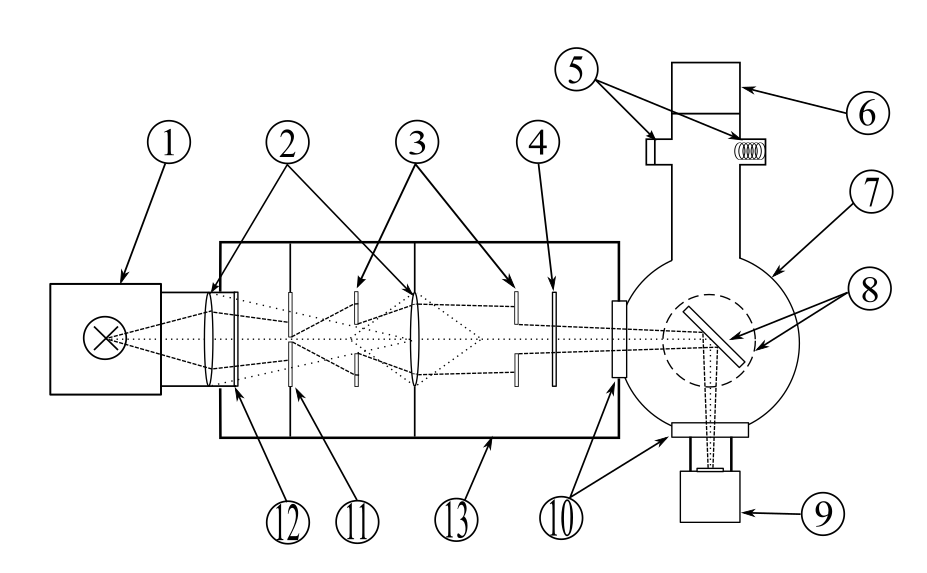}
\caption{Schematic drawing of the setup. 1: Halogen lamp, 2: Lenses, 3: Apertures, 4: Filter, 5: Pressure sensors, 6: Turbo molecular pumping system, 
7: Vacuum chamber, 
8: Foils on cold finger, 
9: Detector (CCD camera), 
10: Quartz glass windows, 11: Pinhole, 
12: Diffusion disc, 
13: Dark box} 
\label{Setup}
\end{figure} 
A schematic drawing of the setup is presented in figure \ref{Setup}. The light source used in the experiment is a $\unit[100]{W}$ halogen lamp (1) manufactured by LOT Quantum Design \cite{LOT} which is cooled by convection. With halogen lamps nearly constant light intensities can be achieved over a long time whereas the intensity of, e.g., LEDs can vary within a few percent.\\
With a collimating lens the source is imaged into a second lens (2). A diffusion disc (12) is mounted to homogenize the light. After the diffusion disc a pinhole (11) is mounted as well as two apertures (3) which prevent stray light from reaching the sample and reduce the diameter of the beam. With the second aperture the cross-section of the resulting beam is further reduced to a diameter of approximately $\unit[10]{mm}$. This beam passes an interchangeable filter (4) and is then reflected off the foil sample (8) under an angle of $\unit[90]{^\circ}$ to a CCD detector (9). The diameter of the beam is chosen in such a way that the full spot size is projected onto the CCD detector. This setup essentially projects an image of the pinhole onto the detector. The tungsten halogen lamp acts together with the collimating lens and the diffusion disc as a light source for the pinhole.\\
The detector (9) used is a monochrome CCD camera (ATIK 383l+ \cite{ATIK}) with a spatial resolution of (3362 x 2504) pixel of the 8.4 Megapixel Kodak KAF 8300 sensor. Each pixel has a size of $(\unit[5.4]{}$ x $\unit[5.4)]{\mu m^2}$ resulting in a sensor size of $(\unit[17.6]{}$ x $\unit[13.5)]{mm^2}$. The camera is using a 16bit Analog to Digital Converter (ADC). To reduce thermal noise the sensor can be cooled to a temperature which is $\sim\Delta$T $=$ $\unit[-40]{^\circ C}$ lower than the ambient air temperature by means of a Peltier cooler. \\
A mechanical shutter is fitted to the camera. This, however, requires a minimum advisable exposure time due to the time in which the shutter is opening and closing. For the ATIK this minimum exposure time is $\unit[200]{ms}$. The camera is linear in pixel value as a function of the exposure time over its full range. Therefore, it is necessary to limit the exposure time to guarantee pixel values below the maximum of around 65000 channels. \\
All optical components are enclosed in a dark box (13). To reduce reflections from scattered light, all walls of this box are coated with an anti-reflective paint (Nextel Velvet-Coating 811-21, produced by 3M).\\
The foil of interest is mounted in a vacuum chamber (7) on a cold finger of a cryocooler (``Cryodrive 3.0'' manufactured by Edwards, UK). The vacuum chamber is needed in order to avoid conductive heat transfer between the wall of the chamber and the cold finger of the cryocooler since this would lead to a high thermal load and therefore, a increased base temperature. Furthermore, it is needed to prevent the condensation of water on the sample. Due to expansion of helium gas in the coldhead it is possible to cool the coldhead down to $\sim\unit[20]{K}$. The chamber is evacuated to a residual pressure $\lesssim\unit[10^{-6}]{mbar}$ by means of a turbomolecular pump (6). This avoids water condensing on the foil and reduces convective heat transfer with the chamber walls.\\
The temperature of the cold finger is measured using a PT100 resistive thermometer. The resistance is read out via a four-point measurement (AVS 47A resistance bridge by Picowatt, Finland). Because PT100 thermometers are not calibrated for temperatures below $\unit[100]{K}$, the resistance of the PT100 was calibrated against a known reference resistance \footnote{The temperature-calibrated reference resistor was obtained from Walther-Meissner-Institut for Low Temperature Research, Garching, Germany.}. During the measurements a temperature of less than $\unit[20]{K}$ was achieved.\\ 
In figure \ref{Holder} the unmounted holder for the foil is shown. A small amount of vacuum grease is put onto a copper plate (1). This allows for a movement of the foil during the cool-down process due to the different coefficients of thermal expansion of copper and the dielectric foil. The foil sample (2) is fixed in between the copper plate and a brass frame (3) which is also prepared with vacuum grease. Four screws fix all the parts. 

\section{Data Analysis and Results}
\subsection{Measurements and Data Analysis}
\label{Ana}

Two types of measurements were performed:\\ 
1) To study the stability of the halogen lamp, a series of single images with a frequency of $\unit[12]{images/hour}$ was recorded. In order to suppress the influence of disturbing stray light and electronic noise a dark image (i.e., an image without light) is subtracted pixel wise from all subsequent images. The mean value over all pixels is then determined to monitor changes in intensity as a function of time. This method has been used with a fixed exposure time of $\unit[7]{sec}$ to check the stability of the measurement over a longer period. \\
2) The relative reflectivity measurements were performed with so-called multi-images where a series of images is combined to one picture. The mean intensities over all pixels at $\unit[300]{K}$ and at $\unit[20]{K}$ are compared. The exposure time is set to a fixed value between $\unit[7]{sec}$ and $\unit[15]{sec}$. An averaged dark picture (also obtained from a multi-measurement with again a set of 50 exposures) is subtracted pixel-wise for both the $\unit[300]{K}$ and the low temperature images. \\
Multi-images are used in order to reduce the effect of statistical fluctuations in the pixel intensities. For a multi-image, all considered images are combined to one by averaging every pixel value of all images. The intensity value in each pixel $x_{i,j}$ is the mean of all values in the pixels $x^l_{i,j}$ where $l$ is the $l$-th image of the measurement taken. The corresponding error $\Delta x_{i,j}$ in each pixel $x_{i,j}$ can be calculated as:

\begin{equation}
\Delta x_{i,j}=\frac{t_N}{\sqrt{N}}\sigma_{x_{i,j}}=\frac{t_N}{\sqrt{N}}\sqrt{<x^2_{i,j}> - <x_{i,j}>^2}
\end{equation}

where $N$ is the number of images in the multi-measurement, $\sigma_{x_{i,j}}$ is the standard deviation of the $N$ measured intensities in pixel $(i,j)$, $<x^2_{i,j}>$ is the mean of the squared intensities, $<x_{i,j}>^2$ is the square of the means and $t_N$ is the prefactor of the Student's t-distribution.\footnote{Student's t-distribution is a continuous probability distribution that arises when estimating the mean of a normally distributed population in situations where the sample size is small and the population standard deviation is unknown \cite{Bevington}.}\\
For a series of measurements where single images are used to measure, e.g., the stability of the signal with time, statistical fluctuations and, therefore, statistical errors can not be taken into account. The error is, in this case, the $1 \sigma$ standard deviation of the dark-noise intensity distribution of the dark image.\\
Further analysis of the data is the same for both types of measurements. The intensities of all pixels $x_{i,j}$ are summed up to obtain a total intensity $x_{tot}$ and divided by the number of pixels. So-called hot pixels (i.e, pixels which do not work properly and show a very high intensity value) are not included in the calculations. The total intensity is:

\begin{equation}
x_{tot}=\frac{1}{n_x n_y}\sum_{i=1}^{n_{x}}\sum_{j=1}^{n_{y}}x_{i,j}
\label{xtot}
\end{equation}

where $n_x$ and $n_y$ are the number of pixels in $x$ and in $y$ direction, respectively. The errors $\Delta x_{i,j}$ for each pixel are summed quadratically since the errors are assumed to be independent. This leads to a total error $\Delta x_{tot}$ for the total intensity:

\begin{equation}
\Delta x_{tot}=\left[ \frac{1}{n_x n_y}\sum_{i=1}^{n_{x}}\sum_{j=1}^{n_{y}}\Delta x_{i,j}^2\right] ^{1/2}
\end{equation}

The total intensities at $\unit[300]{K}$ $\left(x_{tot}^W\right)$ and at $\unit[20]{K}$ $\left(x_{tot}^C\right)$ are compared by calculating the ratio 

\begin{equation}
R = \frac{x_{tot}^C}{x_{tot}^W}=1+\delta R
\label{R}
\end{equation}

where $\delta R$ is an expression for the change of the relative reflectivity. For $\delta R>0$ the intensity at $\unit[20]{K}$ is higher and for $\delta R<0$ lower compared to that at $\unit[300]{K}$. The statistical error is:

\begin{equation}
\Delta R_{stat} = \sqrt{{ \left( {\frac{\Delta x_{tot}^C }{x_{tot}^C}}\right)} ^2+{ \left( {\frac{\Delta x_{tot}^W}{x_{tot}^W}} \right) }^2}
\label{DeltaR}
\end{equation}

The relative deviation $\delta x_{tot}^i$ from the mean of the $i$-th measurement in a series of single-measurements is given by:

\begin{equation}
\delta x_{tot}^i=\frac{x_{tot}^i-\bar x}{\bar x}
\end{equation}

where $\bar x$ is the mean value of all $x_{tot}^i$ in the measurement series.

\begin{figure} [h]
\centering
    \subfigure{\includegraphics[width=0.35\textwidth]{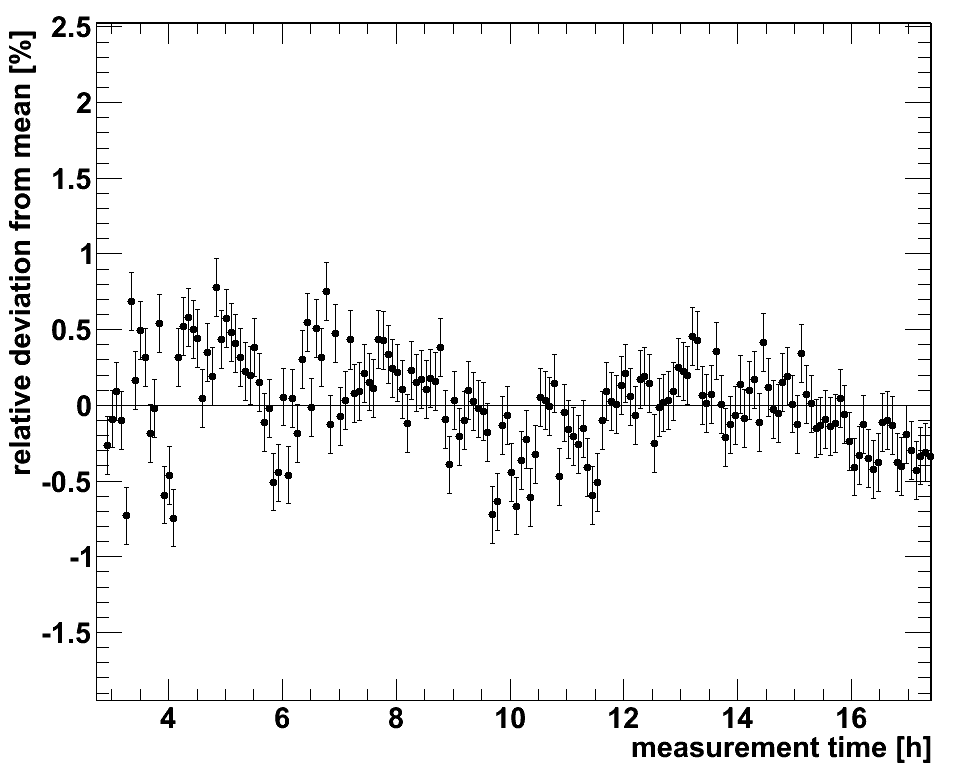}} 
    \vfill
    \subfigure{\includegraphics[width=0.35\textwidth]{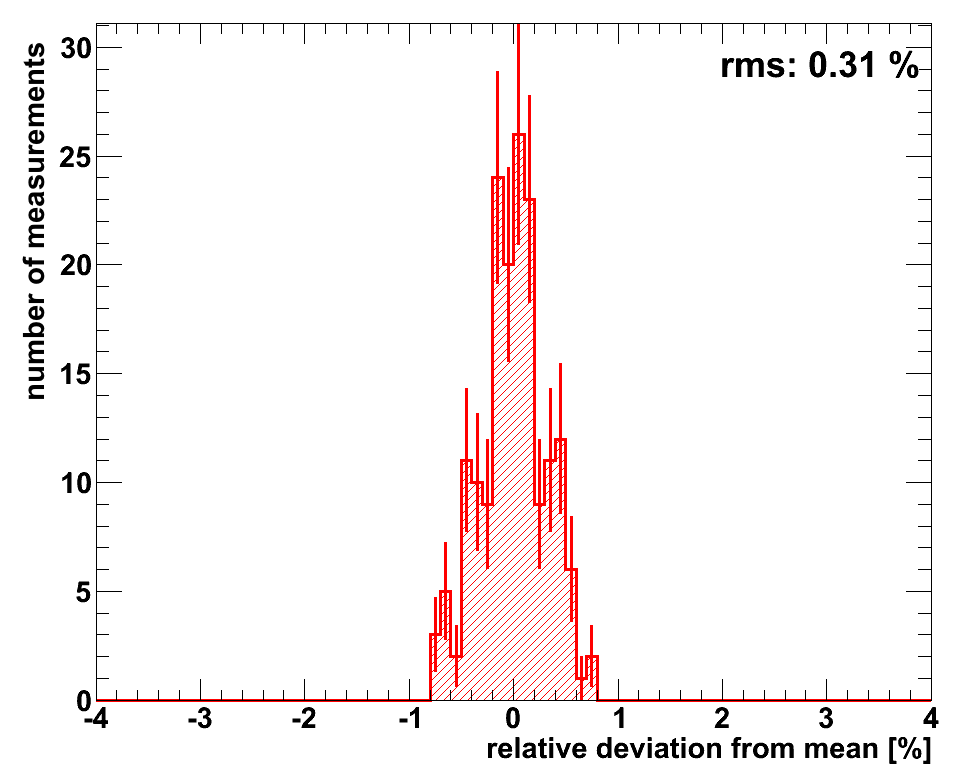}} 
\caption{Stability measurement of the setup. The measurement was started $\unit[3]{h}$ after the warm-up phase. Upper panel: Relative deviation from the mean as a function of time over a period of $\unit[14]{h}$. The scatter of the lamp intensity is smaller than $\unit[1]{\%}$ for all data points. Lower panel: Histogram of the relative deviation from the mean shown in the upper panel. The root mean square of the relative deviation is calculated as $\unit[0.31]{\%}$ which is taken as systematic error of $x_{tot}^C$ and $x_{tot}^W$.} 
\label{RMS}
\end{figure}

\subsection{Stability Measurement}

The systematic uncertainty is mainly due to the stability of the lamp and the optics of the setup. Therefore, a long-time measurement over approximately $\unit[17]{h}$ was performed. First, the lamp was switched on and every 5 minutes an image of the reflected light was taken. The exposure time for the measurement was $\unit[7]{sec}$ and the CCD detector was operated at a temperature of T $=$ $\unit[-15]{^\circ C}$. \\
Figure \ref{RMS} shows the stability measurement after $\unit[3]{h}$ of warm-up phase which is needed to reach thermal equilibrium by convection and radiation. The upper panel shows the relative deviation from the mean as a function of time over a period of $\unit[14]{h}$. The scatter of the intensity is less than $\unit[1]{\%}$ around the mean. In the lower panel the distribution of the relative deviation from the mean is shown. The root mean square of the measurement is $\unit[0.31]{\%}$. Since $x_{tot}^C$ and $x_{tot}^W$ are independent of each other both errors are summed quadratically. This leads to a total systematic error of $\Delta R_{syst}= \sqrt{2}\cdot\unit[0.31]{\%}=\unit[0.44]{\%}$ of the experiment.

\subsection{Relative Change in Reflectivity}

The halogen lamp was switched on for at least $\unit[3]{h}$ before each measurement to guarantee stable operation conditions. For each measurement series 50 images of the reflected beam spot were taken. All samples were measured with the full spectrum of the halogen lamp and with two bandpass filters. The bandpass filters have a center wavelength of $\unit[(420\pm10)]{nm}$ and $\unit[(500\pm50)]{nm}$, respectively \cite{Edmund}.\\
\begin{figure} [h]
\centering
    \subfigure{\includegraphics[width=0.5\textwidth]{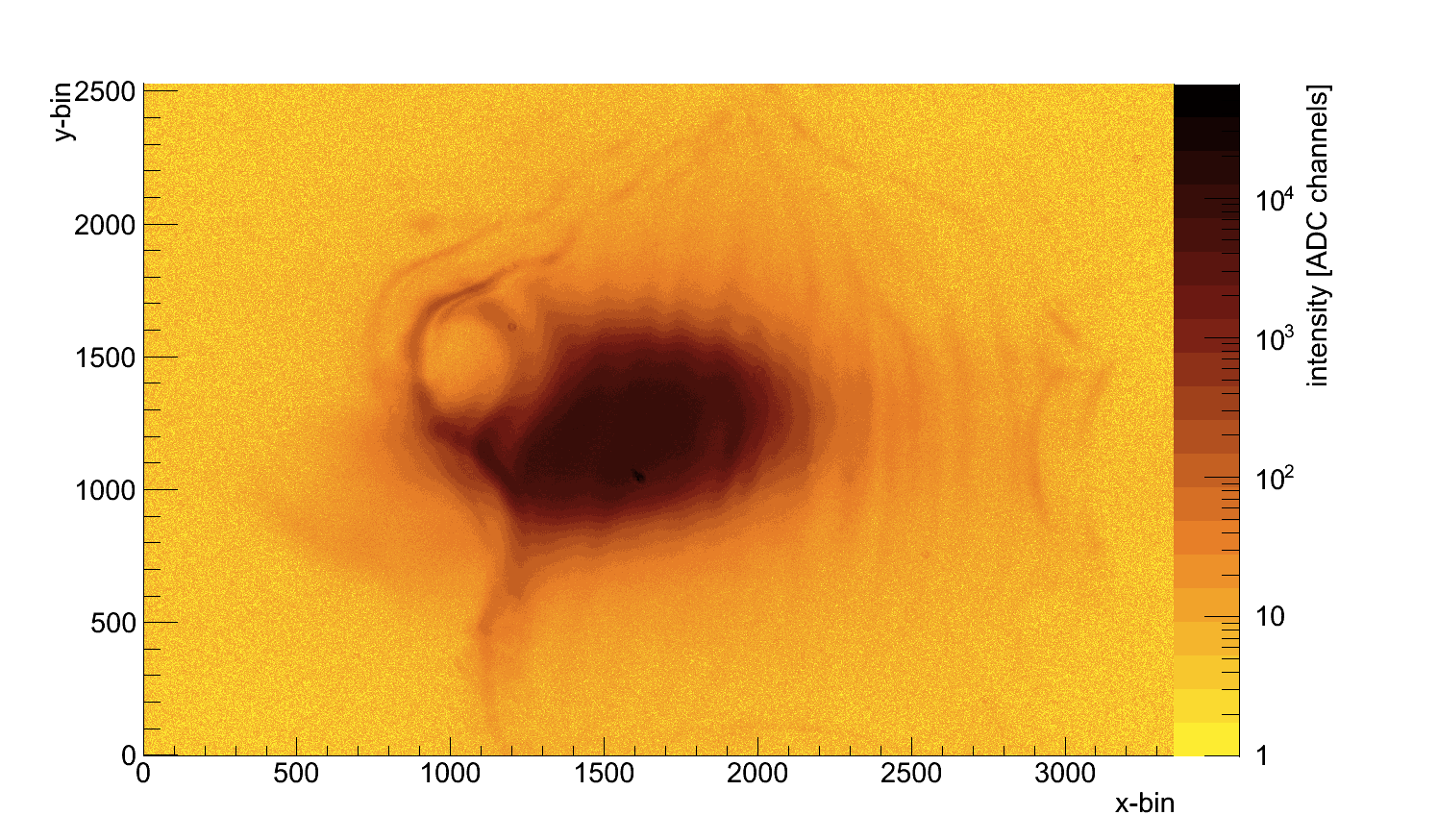}}
    \subfigure{\includegraphics[width=0.5\textwidth]{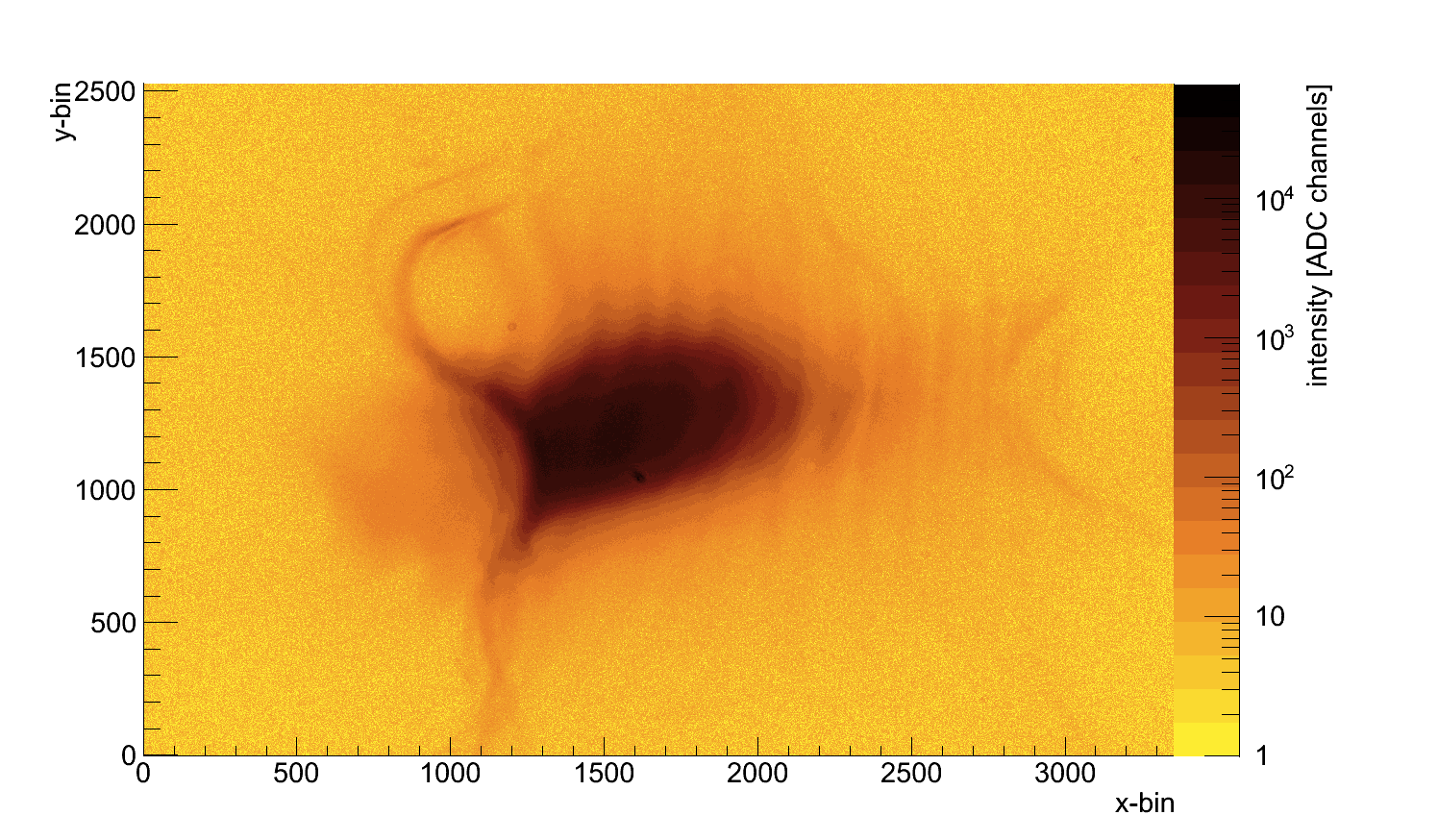}}
\caption{Images of the ``Vikuiti'' front side measured with the full spectrum of the halogen lamp and an exposure time of {$\unit[7]{sec}$}. The CCD detector had a temperature of T $=$ $\unit[-15]{^\circ C}$. The scaling of the intensity (right scale) is logarithmic. Each bin-pair $x_i$ and $y_j$ corresponds to the pixel $ij$. In the upper panel the measurements at $\unit[300]{K}$, in the lower panel at $\unit[20]{K}$ are shown. The spot size (diameter and shape) changes, nevertheless the whole reflected light-beam is detected. This shows that the reflection of the sample changes with temperature but not the integrated light signal.}
\label{Vikallfront}
\end{figure} 
\renewcommand{\arraystretch}{1.5}
\begin{table}[ht]
\small
\begin{center}
\begin{tabular}{ c| c c}
Vikuiti Front & $x_{tot}^C$ [channels] & $x_{tot}^W$ [channels] \\
 \hline
 $\unit[420]{nm}$ & $71.18\pm0.10$ & $71.64\pm0.11$ \\
 $\unit[500]{nm}$ & $668.91\pm1.03$ & $670.74\pm1.18$ \\
 Full Spectrum & $327.89\pm0.48$ & $329.24\pm0.56$ \\
 \hline
 \hline
 Vikuiti Back & $x_{tot}^C$ [channels] & $x_{tot}^W$ [channels]\\
 \hline
 $\unit[420]{nm}$ & $68.69\pm0.12$ & $69.01\pm0.14$\\
 $\unit[500]{nm}$ & $453.41\pm0.88$ & $455.46\pm1.04$ \\
 Full Spectrum & $340.51\pm0.64$ & $343.49\pm0.50$ \\
\end{tabular}
\end{center}
\caption{Total intensities $x_{tot}^C$ and $x_{tot}^W$ (s. eq. \ref{xtot}) for both sides of the ``Vikuiti'' foil. The measurements were performed with two different bandpass filters ($\unit[420]{nm}$, $\unit[500]{nm}$) as well as with the full spectrum of the halogen lamp.}
\label{results1}
\end{table}
The exposure time (between 7 and $\unit[15]{sec}$) was chosen individually for each measurement series, depending on the filter and the size of the reflected spot. The detector was operated at a temperature of T $=$ $\unit[-15]{^\circ C}$. Due to the read-out time of approximately $\unit[10]{sec}$ per image, the complete series was recorded within 14 to $\unit[21]{min}$. \\
Afterwards, the sample was cooled to a temperature of $T\sim\unit[20]{K}$. Once temperature stability was reached, another series of 50 images was recorded with the same parameters. Between the two measurements the lamp was not switched off to avoid changes in intensity of the lamp.\\ 
To measure the background induced by stray light and the electric noise of the camera the halogen lamp was switched off and again 50 images were taken. Also here the exposure time remained the same.\\
\begin{figure} [tbp]
\centering
\includegraphics[width=0.5\textwidth]{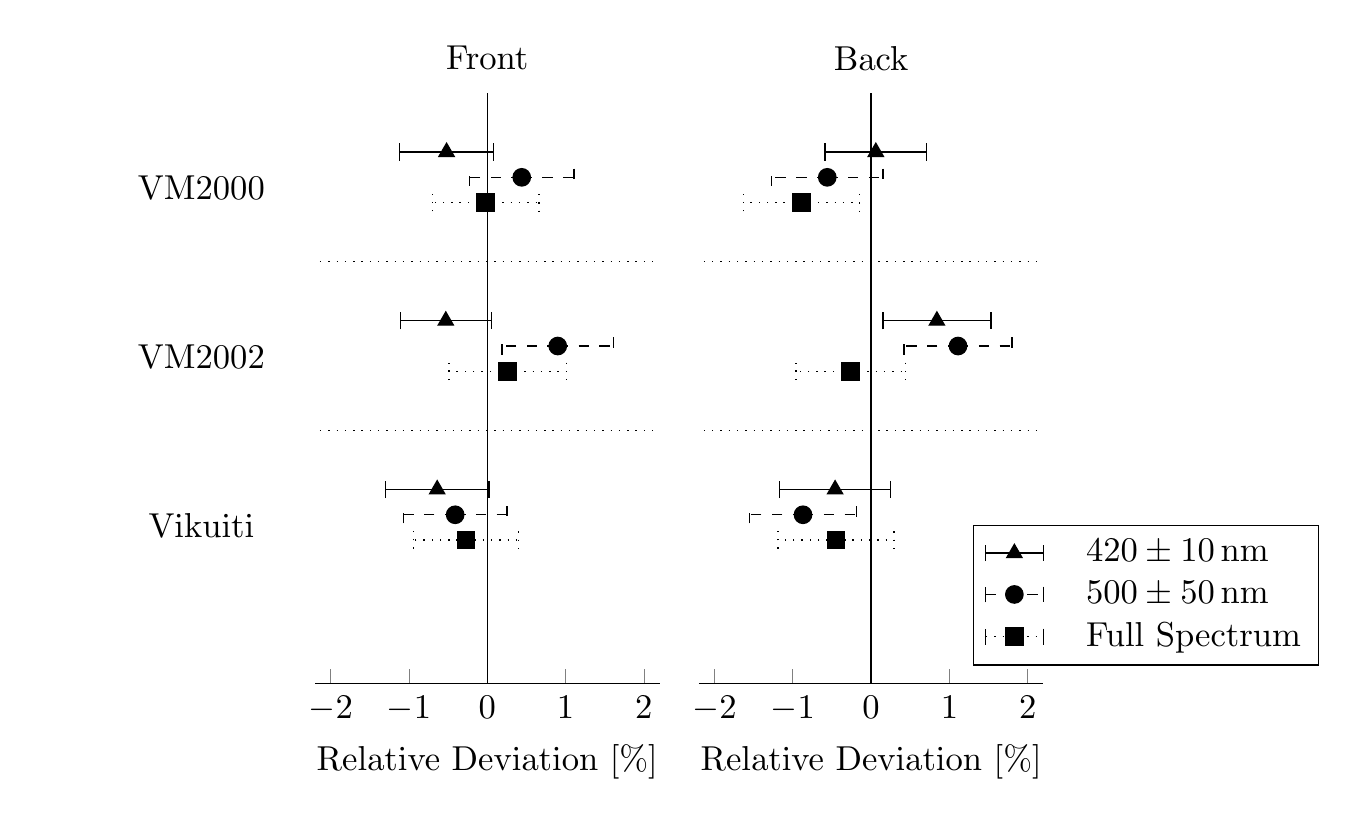}
\caption{Relative deviations from a cold to warm ratio of the reflectivity of $R=1$ (corresponding to the change in reflectivity) are shown. The error bars are the linear sum of systematic and statistical $1\sigma$ errors. Two bandpass filters ($\unit[420]{nm}$, $\unit[500]{nm}$) as well as the full spectrum of the halogen lamp were used. For all foils investigated no significant effect of the temperature on the reflectivity behavior could be found.} 
\label{RelaRef}
\end{figure} 
For an exemplary elaboration of the measurements performed the analysis of the  ``Vikuiti'' front side will be presented. Two images recorded with the full spectrum of the halogen lamp for reflection at the front side of the ``Vikuiti'' are shown in figure \ref{Vikallfront}. The images demonstrate the advantage of the measurement-method applied. The size and the intensity distribution change from $\unit[300]{K}$ (upper panel in fig. \ref{Vikallfront}) to $\unit[20]{K}$ (lower panel in fig. \ref{Vikallfront}) but in both cases the whole spot of the reflected light beam is detected. The total intensities for the ``Vikuiti'' foil are given in table \ref{results1}. For the ``Vikuiti'' front side a cold to warm ratio (s. eqs. \ref{R} and \ref{DeltaR}) of $R= 0.9959\pm0.0022 (\text{stat.})\pm0.0044 (\text{syst.})$ and a relative reflectivity change of $\delta R = [-0.41$ $\pm 0.22$ (stat.) $\pm 0.44$ (syst.)$]\%$ was calculated with the full spectrum of the halogen lamp.\\
In figure \ref{RelaRef} the relative deviations from a cold to warm ratio of $R=1$ are shown for all foils investigated (this deviation corresponds to the change in reflectivity). The error bars shown are the linear sum of the systematic and statistical $1\sigma$ errors. The triangular and circular markers represent the measurements with a $\unit[420]{nm}$ filter and with a $\unit[500]{nm}$ filter, respectively. For the measurement indicated by a quadratic marker no bandpass filter has been used. The measurements performed in this work show no variation of the reflectivity with temperature at a level of $\sim\unit[1]{\%}$. The calculated changes $\delta R$ of the cold to warm ratios (s. eq. \ref{R}) of all foils measured (front side (F) and back side (B)) are given in table \ref{results}.
\begin{table}[h]
\small
\begin{center}
\begin{tabular}{c|c c c }
 & $\unit[420]{nm}$ & $\unit[500]{nm}$ & Full Spectrum\\
\hline 
{$\delta$R$_\text{VM2002,F}$} & {$(-0.53 \pm 0.58)\%$} & {$(+0.26 \pm 0.75)\%$} & {$(+0.90 \pm 0.71)\%$}\\
$\delta$R$_\text{VM2002,B}$ & $(+0.84 \pm 0.69)\%$ & $(-0.26 \pm 0.70)\%$ & $(+1.11 \pm 0.69)\%$\\
$\delta$R$_\text{Vikuiti,F}$ & $(-0.64 \pm 0.66)\%$ & $(-0.27 \pm 0.67)\%$ & $(-0.41 \pm 0.66)\%$ \\
$\delta$R$_\text{Vikuiti,B}$ & $(-0.46 \pm 0.71)\%$ & $(-0.45 \pm 0.74)\%$ & $(-0.87 \pm 0.68)\%$ \\
$\delta$R$_\text{VM2000,F}$ & $(-0.52 \pm 0.60)\%$ & $(-0.02 \pm 0.68)\%$ & $(+0.44 \pm 0.67)\%$ \\
$\delta$R$_\text{VM2000,B}$ & $(+0.06 \pm 0.65)\%$ & $(-0.89 \pm 0.74)\%$ & $(-0.56 \pm 0.71)\%$ \\
\end{tabular}
\end{center}
\caption{Changes $\delta R$ of the cold to warm ratios of all foils investigated (front side (F) and back side (B)). A plus (minus) sign means that the intensity is higher (lower) at $\unit[20]{K}$. For the measurements two bandpass filters were used ($\unit[420]{nm}$, $\unit[500]{nm}$) as well as the full spectrum of the halogen lamp. The measurements show that the temperature has no significant effect on the reflectivity of the foils.}
\label{results}
\end{table}

\section{Summary and Conclusion}

In this work we have studied the relative reflectivity behavior of three commercial multilayer polymer foils (``VM2000'', ``VM2002'', ``Vikuiti'') at $\unit[300]{K}$ and at $\unit[20]{K}$. Using a halogen lamp we generated a light beam which was reflected off the foil sample. Measurements were performed with two different bandpass filters ($\unit[420]{nm}$, $\unit[500]{nm}$) as well as with the full spectrum of the halogen lamp. The 2D intensity profile of the reflected light beam was measured using a CCD camera. The read out of the complete detector area has the advantage of being not sensitive to changes of the reflection due to thermal contraction of the holding structure and foil and, therefore, of the reflected beam diameter. By calculating the ratio of the intensities at $\unit[300]{K}$ and at $\unit[20]{K}$ the relative change in reflectivity was determined. The measurements performed in this work have shown no significant change in the reflectivity with temperature for all foils investigated. This is important for experiments in the field of dark matter and neutrinoless double beta decay (0$\nu\beta\beta$).

\section{Acknowledgements}

This research project was supported by the DFG Cluster of Excellence 'Origin and Structure of the Universe' \linebreak(www.universe-cluster.de) and the Maier-Leibnitz-Laboratory (Garching). We would like to thank Dr. Federica Petricca for providing foil samples.

\bibliographystyle{elsarticle-num-names}
\bibliography{Bibliography}
\end{document}